\documentstyle[amssymbols]{article}

\def\CP{{\Bbb P}}
\def\Cx{{\Bbb C}}
\def\acapo{{\par\noindent }}
\def\Zet{{\Bbb Z}}
\def\frc{{\sim\hskip -2.5pt\sim\hskip -2.5pt\sim\hskip -4.2pt\triangleright }}
\def\mapright#1{\smash{\mathop{\longrightarrow}\limits^{#1}}}

\def\dsl{\displaystyle}
\def\G{\Gamma}
\newtheorem{theorem}{ Proposition} [section]
\newtheorem{prop}{Proposition}[section]
\newtheorem{lem}{ Lemma }[section]
\newtheorem{defi}{Definition}[section]
\date{July~1992}

\title{Singularity, complexity, and quasi--integrability of rational mappings}
\author{G. Falqui\thanks{Supported in part by Minist\`ere de la Recherche et de
la Technologie} ~ and C.-M. Viallet}

\begin{document}
\bibliographystyle{/users/lpthe/viallet/formats/perso}

\begin{titlepage}
\renewcommand{\thepage}{}
\maketitle
\hoffset 5truemm
\begin{center}
Laboratoire de Physique Th\'eorique et des Hautes Energies \\
Unit\'e Associ\'ee au  CNRS (URA 280) \\
Universit\'e Paris VI Bo\^{\i}te 126 \\
Tour 16 -- $1^{\rm er}$ Etage / 4 Place Jussieu \\
75252 PARIS CEDEX 05
\end{center}
\hoffset 0truemm
\vskip 2truecm
\begin{abstract}
 We investigate  global properties of the mappings entering the description of
symmetries of integrable  spin and vertex  models, by exploiting their nature
of birational transformations of projective spaces.
We give an algorithmic analysis of the structure of invariants of such
mappings.
We  discuss some characteristic conditions for their (quasi)--integrability,
and  in particular its  links with their singularities (in the 2--plane).
Finally, we describe  some of their properties {\it qua\/} dynamical systems,
making contact with Arnol'd's notion of complexity, and exemplify remarkable
behaviours.
\end{abstract}
\vfill
PAR--LPTHE 92/26 \hfill To appear in Communications in Mathematical Physics
\end{titlepage}

\section{Introduction}
We want to analyze in detail some realizations of
Coxeter groups~\cite{CoMo65,Hu90}  by birational transformations of projective
spaces which  have been shown to appear in the description of the symmetries of
quantum integrable
systems~\cite{Ba81,Ba82,Fa82}~\cite{bmv1,bmv1b,bmv2,bmv2b,prl}.

The first motivation to look at these realizations resides of course in their
relations with the star--triangle and the Yang--Baxter equations or their
higher dimensional generalizations such as the tetrahedron equations.
A characteristic feature of the orbits of  the known solutions of the
Yang-Baxter equations under these groups is  that they are  confined to
subvarieties of high codimension of the parameter space (actually curves), {\em
signaling the existence of an  unexpectedly large number of algebraically
independent invariants\/}.
The discovery and the analysis of the possible invariants is a decisive step in
the study of the Yang--Baxter (tetrahedron,...)  equations, in particular for
what concerns the so--called baxterization problem~\cite{Jo90}.

Another motivation  is to use these realizations to construct  discrete time
evolution maps, as it is usual in the study {\it a la\/} Poincar\'e of
dynamical systems, by iterating some element of the group.
One of the main   questions in this setting is again to  bring to the light the
possible  presence of invariants and invariant
tori~\cite{henri92,henri52,Wi81}~\cite{PaNiCa90,QuRoTh88,QuRoTh89,Ra91,Ra91b}~\cite{bmv1,bmv1b,bmv2,bmv2b,prl}.

If a realization admits algebraic invariants, we will say it has a property
of quasi--integrability.

In~\cite{bmv1,bmv1b}  it was shown, among other things how the direct graphical
investigation of the group action by means of numerical calculations (``drawing
the picture'')  leads to a nice representation of the orbits.
We want to elaborate on the problem and discuss   the deeper structure of the
non linear realizations by  birational transformations of projective spaces.

We first recall which kind of infinite Coxeter groups arise in the theory of
solvable models of statistical mechanics and  collect  some facts about
birational maps which we shall  use in the sequel.

We then analyze  the cohomological structure of the possible invariants.
We  give the general (albeit formal) solution to the problem in terms of an
algorithmic research of  linear systems on $\CP_n$ satisfying a covariance
property with respect to the realizations we handle.
 We then specialize to the case $n=2$,  analyze the singular locus of the
group, and discuss some illustrative examples.

In the final section we make contact with Arnol'd's notion of
complexity~\cite{Ar90,Ve92}, which measures  the growth of the topological
non--triviality of the intersection  of a fixed subvariety   and the image of
another one  under an iteration map.
We will argue how our notion of quasi--integrability is related to a {\em
polynomial \/} growth of the complexity while the generic one is exponential.

\section{Coxeter groups and birational realizations}

One of the outcomes of~\cite{bmv1}-\cite{prl} is the construction of a number
of groups generated by involutions and of various  rational realizations on
projective spaces.

Consider the  Coxeter group $G$  engendered by $\nu$  involutions $I_1, I_2,
\dots, I_k, \; (k=1\dots \nu)$, {\em verifying no relations other than the
involution property}.
The group $G$ is infinite and there are two essentially different situations.

If $\nu=2$, the group is the infinite dihedral group ${\Bbb Z}_2 \ltimes {\Bbb
Z}$, and all elements may uniquely be written $I_1^\alpha(I_1I_2)^q$, with
$\alpha=0,1$ and $q\in {\Bbb Z}$. The number of elements of given lenght $l$ is
2.

If $\nu \geq 3$, the number of elements of lenght $l$ grows exponentially with
$l$, and the group is in a sense bigger (still countable).

As an example for the groups described  in~\cite{bmv1}-\cite{prl},  the number
$\nu$ of generators depends on the dimension $d$ of the lattice: it is just
$2^{d-1}$ so that if $d=2$,  $G$ is generated by two involutions  and   if
$d\ge 3$, $G$ is generated by more than three involutions.

One may then construct  various  realizations $\Gamma$ of $G$ by explicit
transformations of some projective space.
They are  obtained by specifying the realization of the generators.
Since it is precisely the realizations that we want to study here, and
especially the problem of the existence of invariants, we will mainly talk
about $\Gamma$ and not $G$, and   use the same  notation $I_k$ for the
generators of $G$ and their representatives in $\G$.

The realizations $\G$ we consider are essentially obtained from operations on
matrices, especially matrix inversions, and transpositions of their entries,
the  matrices being  originally matrices of Boltzmann weigths of statistical
mechanical spin and vertex models on the lattice or $R$--matrices of
2--dimensional field theories. The projective space we consider is just the
space of entries of the matrices up to a common multiplicative factor.

Let us describe here typical examples of such  realizations.

Suppose $m$ is a $q\times q$ matrix.
The ordinary matrix inverse $I$ defines an involutive rational transformation
of $\CP_{q^2-1}$, which reads in homogeneous coordinates:
\begin{equation}	\label{inver}
I: \quad m_{ij} \longrightarrow \mbox{ cofactor of }m_{ij}
\end{equation}
We may  also consider  the element by element inverse (so called Hadamard
inverse):
\begin{equation}	\label{hada}
J: \quad m_{ij}  \longrightarrow 1/m_{ij}
\end{equation}
These two inverses appear in the study of spin models.
Notice that $I^2$=$J^2$=1 and there is no other relation between $I$ and $J$.
In particular $I$ and $J$ do not commute and $\varphi=IJ$ is of infinite order.

It is of course possible to define all kinds of block inverses, the size of the
blocks ranging from  the full matrix size (for $I$) and 1 (for $J$).

One may also mix $I$ with linear transformations.
This  happens already  when $I$ and $J$  defined by (\ref{inver}) and
(\ref{hada}) are collineated, i.e when there exists a linear transformation $C$
such that $I=C^{-1}JC$ (see~\cite{bmv1}).
We may also mix $I$ with permutations of the entries.

Let us describe here the  transpositions of entries which appear in the study
of  vertex models on a $d$--dimensional lattice~\cite{bmv2b,prl}.

Suppose $M$ is a multiindex matrix of size  $q^d \times q^d$ written in the
form $M^{i_1 i_2 \dots i_d}_{j_1 j_2 \dots j_d}$.
There exist  $d$ different partial transpositions $t_1,\, t_2, \dots, t_d$ with
the evident definition:
\begin{eqnarray}
        (t_k \, M)^{i_1 \dots i_k \dots i_d}_{j_1 \dots j_k\dots j_d}=
                M^{i_1  \dots j_k\dots  i_d}_{j_1 \dots i_k \dots j_d} .
\end{eqnarray}
We clearly have a product and an inverse $I$ for these multiindex matrices.
We may define $2^{(d-1)}$ new inversions by:
$$
        I_\kappa = t_{\alpha_1}\dots t_{\alpha_s}\,  I \,
                t_{\alpha_{s+1}}  \dots t_{\alpha_d} =
        t_{\alpha_{s+1}}  \dots t_{\alpha_d} \, I \,
                t_{\alpha_1}\dots t_{\alpha_s}
$$
where $\kappa=(\{ \alpha_1,\dots \alpha_s\},\{ \alpha_{s+1},\dots \alpha_d\})$
is a partition of $\{ 1,\dots, d\}$.

These various  inverses yield involutive (bi)rational mappings of
$\CP_{q^{2d}-1}$.
They are related by collineations, easy to write from the representation of the
partial transpositions.
Note that the  product  of all $t_k$'s is the full transposition $t$ and
commutes with all the inverses.
Such realizations appear in the study of vertex models.

We may further enrich the representations by imposing constraints on the
entries of the matrices, provided the transformations are compatible with these
constraints (see~\cite{bmv1} for the notion of admissible patterns).
This yields realizations on projective spaces of lower dimensions.

Needless to say that along these ideas, one may construct a variety of
involutions acting on various projective spaces.

We stress that, at the level of the  realization, there may exist additional
relations between the generators, possibly making it  finite (as in example
(\ref{fm})).

\section{Some facts about rational mappings} In this section we
collect
some results about rational and birational mappings between algebraic
varieties  (see for example~ \cite{Wa50,Mu78,GrHa78,Sh77}).
\begin{defi}
 A correspondence  $\varphi_Z:X\to Y$ between algebraic varieties $X$ and $Y$
is an algebraic subset  $Z\subset X\times Y$.
$\varphi_Z$ is  a rational map if there is a Zariski open set $U\subset X$ on
which the correspondence is one to one.
\end{defi}
If $Z\subset X\times Y$ is a rational map,  the inverse correspondence is
defined  by the  graph $Z^{-1}:=\{(y,x)\in Y\times X\ |(x,y)\in Z\}$.
 If the correspondences $Z$ and $Z^{-1}$ are both
rational mappings, then $Z$ (or  $\varphi_Z$) is called a
birational transformation.
A  birational map is a  biholomorphism except on subvarieties of codimension at
least two.

A linear system $D$ on $X$ is a non empty linear subspace of the space
of global sections of some line bundle ${\cal L}$ over $X$.
Its base locus is the set  of common zeroes of all sections in $D$.
A remarkable result is that \cite{Mu78} there exists a one to one
correspondence between linear systems on $X$ of  dimension $d$  with base locus
of codimension not less that 2  and rational maps $X\mapright{\phi}\CP_{d-1}$
up to  projective automorphisms of $\CP_{d-1}$.

For what we are concerned with, the paradigm  of  rational map is the
so--called $\sigma$--process or blow up of a point.

We refer to~\cite{GrHa78,Sh77} for the general definitions.
We shall call Hadamard inversion and generically denote by $J$ the prototypical
birational mapping in  $\CP_n$.
Let $x_0,x_1,\dots, x_n$ and $y_0,y_1,\dots, y_n$ be coordinates in two
different copies of $\CP_n$ and let us consider the algebraic set
$Z\subset\CP_n\times\CP_n$ given by the $n$ equations $x_0 y_0 = x_1 y_1
=\dots= x_n y_n$.
By definition, the graph of $J$ is $Z$.

It is valuable to specialize to $n=2$: outside the triangle $x_0x_1x_2=0$, $Z$
is the graph of the map $[x_0,x_1,x_2]\frc[1/x_0,1/x_1,1/x_2]$.
The generic point on the line $x_i=0$ is sent into the point $p_i$ whose only
non--vanishing coordinate is $y_i$, while to the points $p_i$  corresponds the
entire line $y_i=0$.
One can say that $J$ blows up $p_i$ to the line $x_i=0$ and blows down the line
$x_i=0$ to the point $p_i$.

Finally, we recall  the following  properties, which will be used in what
follows.
{}From the description of a  birational mapping as an algebraic set in the
product $X\times Y$ it is apparent that $\varphi$ blows up $p$ to a divisor $D$
if and only if $\varphi^{-1}$ blows down $D$ to $p$, and it is also evident
that different points $p_1$ and $p_2$ cannot be blown up to the same divisor
$D$ (otherwise  the inverse map would not be rational).

Also, it is a standard result that blowing up a point adds a free factor $\Zet$
in the Picard group of $X$.
It follows that  if $\varphi$ is a birational map in $\CP_2$ which blows up $n$
points, then it must blow down exactly $n$ exceptional divisors,  as explained
above.

\section{Invariants and quasi--integrability}
Let $\Gamma$ be a  group of birational transformations in $\CP_n$.
A meromorphic function $\Delta:\CP_n\to \CP_1$ deserves to be called a
$\Gamma$--invariant if it satisfies
\begin{equation}
\Delta(g(x))=\Delta(x)\quad \forall g \in \G.
\end{equation}
Since a meromorphic function on an algebraic variety can be   thought of as the
ratio of two  sections of a suitable line bundle, we are naturally lead to the
following scheme.

A $\G$ one--cocycle is a collection of sections  $a(g,x)$ of some line bundles
over $\CP_n$ satisfying the cocycle condition:
\begin{equation}
\label{coc}
a(g_1  \;  g_2,x)= a(g_1,g_2 x)\cdot a(g_2,x)
\end{equation}
A section  $\sigma$ of a  line bundle  will be called {\em $a$--covariant} (for
some cocycle a) if   the equation
\begin{equation}
\label{cob}
\sigma(g x)=a(g,x)\cdot\sigma(x)
\end{equation}
holds.
This equation may be reformulated in group cohomology terms~\cite{HoSe53b}  as
 $ a = \delta \; \sigma $  meaning  that $a$ is actually a coboundary.

Finding a $\G$--invariant is equivalent to finding two sections $\sigma_1 $ and
$\sigma_2$ verifying equation (\ref{cob}) for the {\em  same} $a$.
This means  that we are interested in  the coboundaries of the $0$--cochains
 rather than the cohomology groups. The strategy is to find  1-cocycles
admitting a sufficient number of primitives.

 As a side remark, we notice  that these equations, which will play a prominent
role in the sequel, are well defined at all points of $\CP_n$.
In fact for any  birational transformation $g$, the singular locus is a
subvariety of codimension greater than or equal to two in $\CP_n$, and the
equations above admit a unique holomorphic extension to the whole of $\CP_n$,
even if they have meaning in the point set sense only on the nonsingular locus
of $g$.

We are interested in the case where  $\Gamma$ is a Coxeter group, i.e.
is  generated by $\nu$  involutions $I_k$ of degree $d_k$ (the degree is a
natural notion in terms of the homogeneous coordinates   $[x_0,\ldots,x_n]$).
A  $\Gamma$ one--cocycle will be
completely specified by the assignment of $\nu$ sections $a(I_k,x)$.
Remarkably,  the possible values of $a(I_k,x)$ may be found explicitely.

Each involution $I_k$ defines a  characteristic polynomial $\phi_k$ of degree
$d_k^2-1$  in the following manner. The $I_k$ being involutions,  $I_k^2$
appears as the multiplication by a degree $d_k^2-1$ polynomial
$\phi_k(x_0,\ldots,x_n)$. We then have the following
\begin{lem}
If $a(g,x)$ is a trivial
$\Gamma$ cocycle, the sections $a(I_k,x)$ divide a suitable power of
$\phi_i$ for $i=1,\ldots ,\nu$.
\end{lem}
\acapo {\bf Proof.} Suppose $a =\delta \; \sigma$. Then, from  the coboundary
equation (\ref{cob}) for $g=I_i^2$ we get that $a(I^2_i,x)=\phi(x)^m$, with
$m=deg(\sigma)$. The assertion follows from the cocycle condition (\ref{coc}).
\begin{defi}
We shall say that  $\Gamma$ is   {\em collineated\/}  when
its generators are all conjugated by means of elements of $PGL(n+1,\Cx)$
to a standard one, $K$.
\end{defi}

This is the case for a number of models among which the Baxter model ($n=3$)
and the examples of section \ref{examples}.

If  the characteristic polynomial $\phi_K$  factorizes into
$\phi_K=\prod_{l=0}^j p_l^{d_l}$,
then it  is clear that for every $i$ there exist a set of global homogeneous
coordinates $[X_0^{(i)},\ldots,X_n^{(i)}]$ in which $a(I_i,x)$ will be the
product of the same polynomials $p_l$, possibly weighted with different
exponents $d^\prime_l$.

\acapo {\bf Remark}. When $K$ is the Hadamard inversion $J$  in $\CP_n$  a
straightforward computation shows that
$\phi_J= \prod_{l=0}^n x_l^{(n-1)}$ so that we get

\begin{theorem}
A coboundary for a collineated Coxeter group of birational transformations with
generators conjugated to the Hadamard inversion $J$ in $\CP_n$ is a monomial
in some suitable homogeneous coordinates, with coefficient $\pm 1$.
\end{theorem}

It  is apparent that our  cohomological setting gives an algorithmic
prescription for the search for invariants: find first the characteristic
polynomials $\phi_k$, which is straightforward, then the possible coboundaries,
which is a factorization problem, and check how many primitives they have,
which amounts to solving a linear problem.

\bigskip

The realization  $\Gamma$ in  $\CP_n$ admits  $p$ invariants $\Delta_1,\dots
\Delta_p$ if  there exists  an $a$--covariant linear system $L^d_p$ of
projective dimension $p$ and  degree $d$ for some cocycle $a$.
It is not guaranteed that the orbits of a realization admitting $p$ invariants
lie on subvarieties of dimension $n-p$. Indeed the question of the {\em
algebraic \/} independence of the invariants has to be examined
further~\cite{prl2}.

\bigskip

One has to realize how exceptional is  the existence  of any invariant
 of the birational realization, as the following argument shows.

Let us  consider  covariance with respect to  one generator $K$ of $\Gamma$.
Suppose $K$ is of degree  $d_K$,  and  suppose we are looking for an invariant
of degree $m$.
Clearly from equation (\ref{cob})  the degree $q$ of the cocycle is related to
$d_K$ and $m$ by  $m(d_K-1)=q$.
The dimension of the space  ${\cal P}(m,n)$   of homogeneous polynomials of
degree $m$ in $n$ variables  is  $n+m-1\choose m-1$.
The requirement of  $K$--covariance selects a  linear system $L^K$ in   ${\cal
P}(m,n)$, of generic  dimension at most ${1\over 2} {n+m-1\choose m-1}$.
Imposing  the same condition for another generator $I$ leads to look for the
intersection of hyperplanes of at most complementary dimensions in ${\cal
P}(m,n)$. This intersection is generically empty. As a consequence we have
\begin{theorem}
The action of a generic  Coxeter group of birational transformations in $\CP_n$
does  not admit any non--trivial invariant.
\end{theorem}
\acapo
In the particular case of collineated realizations, this leads to the further
property, which we prove for $\nu=2$ for simplicity.
\begin{theorem}
 The set of {\em collineated}   groups admitting a non--trivial invariant has
the structure of  a quasi--projective variety.
\end{theorem}
\acapo{\bf Proof}.
The group $PGL(n+1,\Cx)$ admits a natural structure
of quasi--projective variety, since it is  identified with
the complement of the degree $(n+1)$ hypersurface $det A =0$ in
$\CP^{(n+1)^2-1}$.
Choosing  $K$ as prototypical generator of $\G$,
$I$ is specified (in the collineated case) by the choice of an element $C\in
PGL(n+1,\Cx)$ by  $I=C^{-1}KC$.
If for any polynomial $P$ we define $P_C:=P(C^{-1}x)$ and set  $y=Cx$, then
the covariance equation with respect to  $I$ reads
$$P_C(Ky)=a_I(C^{-1}y)P_C(y)$$
\noindent
For each possible cocycle $a$, this  is an algebraic equation
in the parameters $t_a$ of the matrix $C$.
The coordinates of any basis $\{e_\alpha\}$ in $L^I$ with respect to a basis in
${\cal P}(m,n)$ completed from a basis of $L^K$ may be expressed as algebraic
functions of the $t_a$'s.
The condition of nontriviality of the intersection is a condition on the rank
of the matrix
$$B:=\left[\matrix{
	{\bf 1}				&{\bf 0}	\cr
	e_\alpha^i(t_1,..t_{(n+1)^2-1})	&\dots		\cr}
\right]$$
which in turn is an algebraic condition on the
$e_\alpha^i(t_1,..t_{(n+1)^2-1})$, concluding  the proof.

\section{Realizations in the 2-plane}

\label{cp2}
Let us
discuss mappings in $\CP_2$. Here our analysis is made quite complete
by the fact that all rational maps from an algebraic surface can be
described in terms of $\sigma$ processes  and also that singularities of
birational maps can occur only at points.
Moreover  a theorem of M. Noether~\cite{Sh77} assures that the group of
birational maps is generated by the inversion $J$ together with the projective
group $PGL(3,C)$.
We will restrict ourselves here to $\nu=2$, i.e. $\G$ is generated by two
involutions $I$ and $K$.

\begin{defi}The singular locus of $\G$, $S(\G)$ is
defined as $$ S(\G)=\{x\in\CP_2 | \exists g\in \G\  {\rm s.t. \ } g\
{\rm blows\ } x\ {\rm up\ } \}$$
\end{defi}
\noindent Since $\G$ is generated  by  $I$ and $K$, $S(\G)$ is obtained by the
action of $\G$ on the singular points of $I$ and $K$.

\begin{defi}
We will  say that $\G$  is properly singular if $S(\G)$ contains at least four
points in general position, i.e. such that no three of them are aligned.
\end{defi}

\noindent Let  $\Pi(\G)$ be the set of singular divisors:
$$\Pi(\G)=\{\pi\in Div(\CP_2) |\exists g\in \G\  {\rm s.t. \ } g\
{\rm blows\ }\ \pi \mbox{  down to a  point  }\}$$ Let us suppose
that  $\G$ admits a rational invariant $\Delta$ and let us look at what happens
at points in $S(\G)$.
The equation $\Delta(g(x))= \Delta(x)$ is clearly meaningless at the
indetermination points of $\Delta$.
Noticing that $\forall x\in S(\G)$, $\exists \pi_x\in\Pi(\G)$ such that $\Pi_x$
is blown down to $x$ by some element $g \in \G$,  we can conclude that
$x\in S(\G)$ is either an indetermination point of $\Delta$ or $\Delta$ is
constant along the corresponding divisor $\pi_x$,  i.e. $\pi_x$ belongs to the
pencil of curves $\Delta= const$.
Then we can state the

\begin{prop}
If $S(\G)$ is infinite, and $\G$ is properly singular,  then  $\G$ does not
admit any invariant.
\end{prop}
\acapo{\bf Proof.} With the above
observation  in mind, the only case  we must rule out is that
$\G$ admits an infinite number of singular points and only a finite number of
singular divisors, since by Bezout's theorem, any rational degree $d$ invariant
admits at most $d^2$ indetermination points. Let us suppose that $\#S(\G)=
\infty$ and $\# \Pi(\G)=N$ and let us consider the set
${\cal P}= \{ \Pi_\alpha \}$ $(\alpha = 1 \dots 2^N)$  of  parts  of $\Pi(\G)$.
We can make a partition of $\G$ into $2^N$ disjoint subset
$\G=\bigcup_{\alpha=1}^N \G_\alpha$,  where
$$\G_\alpha=\{g\in \G \vert  g \mbox{ blows down exactly all the divisors in }
\Pi_\alpha\}$$
\noindent
At least one of the  $\G_\alpha$, say $\G_0$ is infinite.
If we then consider any pair of elements  $(h_0, l_0)$ in
$\G_0$,   the product $h_0 \cdot l_0^{-1}$ is a birational map without
singularities, and hence an element $D\in PGL(3,\Cx)$.
Since  $\G_0$ is infinite, we can arrange things so that $D$ is a non trivial
word  of even lenght.
Since  words of odd length in $\G$ are involutions one has:
\begin{equation}
\nonumber
D K D = K \qquad D  I D =  I
\end{equation}
 It follows that $D$ permutes the points of $S(\G)$, and some  power $D^k$ must
be  the identity.
As a consequence,  there is some non--trivial product $(IK)^l={\bf 1}$, meaning
that actually $\G$ is  finite, and contradicting the infiniteness of $S(\G)$.
\acapo{\bf Remark.} The above proposition  proves that a necessary condition
for existence of an invariant is the  finiteness of the singular locus.
It is tempting to conjecture that this is also a sufficient condition for any
properly singular realization.
This is unfortunately not the case, as the  example of section (\ref{fdm}) will
show.

\bigskip

The next step is to try to relate the singular
locus of the group with the singular locus of the would--be invariant. Let
the $\Delta$ be a rational invariant for $\G$ and let $S(\Delta)$ be the
set of its indetermination points.
If $y\in S(\Delta)$ and $y\notin S(\G)$, the (finite) orbit $\G_y$ of $y$ is
made out of indetermination points of $\Delta$, on which  every element $g\in
\G$ is regular.
Since $\G_y$  is of even order, say  $2l$, $y$ is a fixed point for
$(IK)^{n\cdot l}, \quad n\in\Zet$.
Let $\tilde\CP_2^y\mapright{pr}\CP_2$ be the 2--plane $\CP_2$ blown up at $y$.
Since we are working with $\CP_2$,  $\Delta$ extends to a function
$\tilde\Delta$ on  $\tilde\CP_2^y$ with no indetermination points in a
neighbourhood of the  exceptional divisor $E=pr^{-1}(y)$.
The restriction $\varphi =  \tilde\Delta\vert_E$ is a meromorphic map on
$\CP_1$ whose value is  determined by the limits of $\Delta(x)$ along lines in
$\CP_2$ passing through $y$.

We will say that  $\G$ is  non--degenerate if  for every $g\in \G$  the tangent
map to $g$ is non--nilpotent at every isolated fixed point of $g$.
Then we can  state the
\begin{prop}
Let $\G$ be a non--degenerate Coxeter group of birational transformations of
$\CP_2$.
If $\G$ admits an invariant $\Delta$, then $S(\Delta)\subset S(\G)$.
\end{prop}	{\bf Proof.} Suppose  $y\in S(\Delta)$ and $y\notin S(\G)$; the
non--degeneracy of $\G$ ensures that there is at least one line in $P
T_{\CP_2}(y)$ whose orbit under the tangents to $(IK)^{n\cdot l},\quad
l\in\Zet$ is infinite (recall that $(IK)^l$ is a minimal length  element in
$\G$ which fixes $y$).
Hence $\varphi$ as defined above assumes the same value on an infinite
number of points, i.e. it is constant over the whole of $E_y$.
But this in turn implies that $\Delta$ is well defined at $y$ which contradicts
$y \in S(\Delta)$ and ends the proof.
\label{cex}

Notice that the converse inclusion  $S(\G)\subset S(\Delta)$ does not hold in
general. If  the inclusion is strict, we have seen that for any $x\in S(\G)$
and not in $S(\Delta)$, the invariant $\Delta$ is constant along some divisor
$\pi_x$. This gives a useful information about the invariant (see example
\ref{at}).

\bigskip

As for what the  singularities of the generic curve $\Sigma_\lambda$ of the
pencil are related  to the singular orbit $S(\G)$ the following considerations
hold.
By Bertini's theorem, generic curves are smooth outside the base locus
$S(\Delta)$ whence the chain of  inclusions
\begin{equation}\label{incl}
S(\Sigma_\lambda)\subset S(\Delta)\subset S(\G)
\end{equation}
Moreover, exploiting the genus formula
for singular curves  one can give relations{~\footnote{We
thank M. Talon for useful remarks on this point.} between the degree $d$ of
the invariant and $S(\G)$.
If $\Sigma$ is a degree $d$ curve in $\CP_2$   with singular locus
$Sing(\Sigma)$,  then \cite{Wa50}
$$
g(\Sigma)={{(d-1)\cdot (d-2)}\over 2}-\sum_{p\in {\rm Sing}(\Sigma)}\delta_p
$$
with $\delta_p$ depending on the type of the singularity at p.
Since  $\Sigma_\lambda$ is irreducible and admits $\G$ as an infinite group of
automorphisms, thanks to the inclusion (\ref{incl}) and to the fact that
$\delta_p = 0$ if $p$ is not in $Sing(\Sigma_\lambda)$, one has
$${{(d-1)\cdot (d-2)}\over 2}-\sum_{p\in S(\G)}\delta_p=\cases{0\cr 1\cr}$$
\noindent
This relation shows that there is a balance between the degree of
the invariants and the number and nature of the singular points of the generic
curve, and consequently of $\G$.

\section{ Examples}
\label{examples}
We  describe  here some specific examples  in $\CP_2$ with two generators $I$
and $J$. The physical origin of the models we will be dealing with
is to be found notably among two--dimensional spin  models with interaction
along the edges~\cite{Ba81}, and this explains the terminology we use.

We fix  $J$ to be the Hadamard inversion $[x_i]\frc [1/x_i]$,
and $I$ to be  $I=C^{-1} J C$ with the collineation matrix
$C\in PGL(3,\Cx)$. We will concentrate on the parametrization by $C$ and
examine some algebraic families in $PGL(3,\Cx)$. For some of them $\Gamma $
admits a non trivial algebraic invariant.

We can associate to $\G$ a diagram  $D_\G$ whose vertices are the points in
$S(\G)$ and where two vertices $p_1$ and $p_2 \in S_\G$ are joined by an edge
if either $p_1=J p_2$ or $p_1=I p_2$. The edges are oriented if $I$ (resp. $J$)
can be applied in one direction only.
The diagram does not characterize the examples, but is a useful tool. In
particular the families we give here have been obtained by deformations of
given collineation,  demanding the stability of the topology of the diagram.

The singular points of $J$ are $P_1=[1,0,0]$, $P_2=[0,1,0]$, $P_3=[0,0,1]$, and
the one of $I$ are $Q_i=C^{-1}(P_i), (i=1,2,3)$.

One should notice that a number of the  families we produce in this way  fall
into the general form found in~\cite{Ja91} (i.e. verifying $C([1,0,0])=[1,1,1]$
and $C^2([1,0,0]=[1,0,0]$), but with {\em  non integer entries}.
This general form depends on four parameters:
\begin{equation}
\label{jaeger}
\left[\matrix{	2&	\alpha&		\beta	\cr
		2&	-1+\gamma&	-1-\gamma	\cr
		2&	-1-\delta&	-1+\delta	\cr}\right]
\end{equation}

\subsection{The $Z_5$ family}
The general $Z_5$ (five-state chiral Potts) model is described by a $5\times 5$
cyclic  matrix.
We  may consider its 2--parameter reduction obtained by imposing that the
matrix is symmetric.
It falls into  a family of quasi--integrable  models parameterized by
a complex number $q$, which (whenever this makes sense)  is identifiable with
the square root  of the number of states~\cite{defo92}.
The collineation matrix is
\begin{equation}
C_{Z_5}(q)=\left[\matrix{
		2 &	{{q^2-1}}	&{{q^2-1}}\cr
		2 & 	-1+q		& -1-q \cr
		2 &	-1-q 		& -1+q \cr	}\right]
\end{equation}
Its  singular set is made of {\em ten} points:$P_1,P_2,P_3$ which are the usual
singular points of $J$,  $Q_1=[1,1,1]$ and $Q_\pm=[-4, {{1\pm {q}}},{{1\mp
{q}}}]$, the singular points of $I$, and $R_2=I(P_2),\>
R_3=I(P_3),\>R_\pm=J(Q_\pm)$.
All these points are  indetermination points for the invariant
$$\Delta_{Z_q}= {{(x-z)(y-z)((q-1)(x^2+y^2)+2(q+1)xy)}\over{(2+(q-2)
(zx+zy)+2xy)(x-y)^2}}$$
Notice that here $S(\Delta)=S(\G)$. The diagram $D_\G$ is the following:
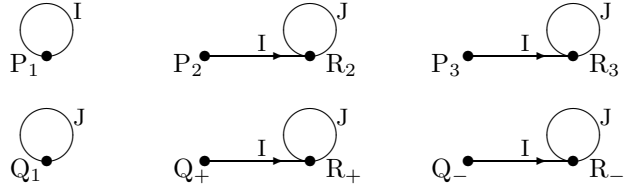
\setlength{\unitlength}{.7mm}
\par\null\hfill
\begin{picture}(140,45)
        \put(20,10){\circle*{2}}\put(20,15){\circle{10}}
        \put(25,17){\small J}\put(13,7){Q$_1$}

        \put(20,30){\circle*{2}}\put(20,35){\circle{10}}
        \put(25,37){\small I}\put(13,27){P$_1$}

\put(70,30){\circle*{2}}\put(70,35){\circle{10}}
\put(75,37){\small J}\put(73,27){R$_2$}

\put(120,30){\circle*{2}}\put(120,35){\circle{10}}
\put(125,37){\small J}\put(123,27){R$_3$}

\put(70,10){\circle*{2}}
\put(70,15){\circle{10}}
\put(73,7){R$_+$}\put(75,17){\small J}

\put(120,10){\circle*{2}}
\put(120,15){\circle{10}}
\put(123,7){R$_-$}\put(125,17){\small J}

\put(50,30){\circle*{2}}\put(44,27){P$_2$}
\put(50,30){\line(1,0){20}}\put(60,31){\small I}
\put(50,30){\vector(1,0){15}}

\put(100,30){\circle*{2}}\put(93,27){P$_3$}
\put(100,30){\line(1,0){20}}\put(110,31){\small I}
\put(100,30){\vector(1,0){15}}
\put(50,10){\circle*{2}}\put(44,7){Q$_+$}
\put(50,10){\line(1,0){20}}\put(60,11){\small I}
\put(50,10){\vector(1,0){15}}
\put(100,10){\circle*{2}}\put(93,7){Q$_-$}
\put(100,10){\line(1,0){20}}\put(110,11){\small I}
\put(100,10){\vector(1,0){15}}
\end{picture}
\hfill\null\par

\subsection{The BMV family}
The matrix of (Boltzmann) weights of the BMV model \cite{bmv1} is
$$
\left[\matrix{
	x&	y&	z&	y&	z&	z\cr
	z&	x&	y&	z&	y&	z\cr
	y&	z&	x&	z&	z&	y\cr
	y&	z&	z&	x&	z&	y\cr
	z&	y&	z&	y&	x&	z\cr
	z&	z&	y&	z&	y&	x\cr
}\right]
$$
The inversion $I$ is just the matrix inversion. This model
pertains to the one--parameter family of quasi--integrable models
whose collineation  matrix is
\begin{equation}
C_{BMV}(w)=\left[\matrix{1&w-1&w\cr 1&-1&0\cr 1&0&-1\cr}\right]
\end{equation}
and is reached for the value $w=3$.
The singular locus is made out  of $P_1,P_2,P_3$ and
$$Q_1=[1,1,1], Q_2=[ w-1, -(w+1),w-1], Q_3=[1,1,-1],$$
\noindent
together with an extra point $R= J(Q_2)=[w^2-1,(w-1)^2,w^2-1 ]$. The case
$w=1$ is singular.
The family admits the invariant
$$\Delta_{BMV}(w)={{P_w^2(x,y,z) Q_w(x,y,z)}
\over{(y+z)^4(x-z)^2(x-y)}}$$ where
$$P_w(x,y,z)=(1-w)(z^2-xy)+(w-3)z(x-y),$$ $$
Q_w(x,y,z)=(1-w^2)(y^3-xz^2)+(w^2-4w-1)y^2(x-z)+2(w-1)^2yz(x-y)$$
 Here again  $S(\Delta)=S(\G)$ and the  singular graph is as follows:

\par\null\hfill
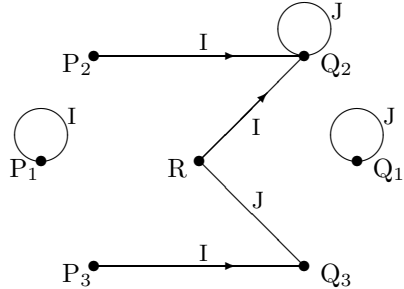
\begin{picture}(80,65)(-10,0)
\put(0,30){\circle*{2}}
\put(0,35){\circle{10}}
\put(5,37){\small I}\put(-6,27){P$_1$}

\put(60,30){\circle*{2}}\put(60,35){\circle{10}}
\put(65,37){\small J}\put(63,27){Q$_1$}

\put(10,50){\circle*{2}}
\put(4,47){P$_2$}

\put(50,50){\circle*{2}}\put(50,55){\circle{10}}
\put(55,57){\small J}\put(53,47){Q$_2$}

\put(10,10){\circle*{2}}
\put(4,7){P$_3$}
\put(50,10){\circle*{2}}
\put(53,7){Q$_3$}
\put(30,30){\circle*{2}}\put(24,27){R}
\put(30,30){\line(1,1){20}}\put(30,30){\vector(1,1){13}}
\put(30,30){\line(1,-1){20}}
\put(10,50){\line(1,0){40}}
\put(10,50){\vector(1,0){27}}
\put(10,10){\line(1,0){40}}\put(10,10){\vector(1,0){27}}\put(30,51){\small I}
\put(30,11){\small I}\put(40,21){\small J}\put(40,35){\small I}
\end{picture}
\hfill\null\par

\subsection{The symmetric Ashkin--Teller model}
\label{at}
By symmetric Ashkin--Teller model we understand
the $4$--state spin model with the cyclic and symmetric matrix of Boltzmann
weights
$$\left[\matrix{
	x_0&	x_1&	x_2&	x_1	\cr
	x_1&	x_0&	x_1&	x_2	\cr
	x_2&	x_1&	x_0&	x_1	\cr
	x_1&	x_2&	x_1&	x_0	\cr
}\right]$$
for which the matrix inversion is collineated
to the Hadamard inversion by means of the matrix
\begin{equation}
C_{AT}=\left[\matrix{
	1&	2&	1	\cr
	1&	0&	-1	\cr
	1&	-2&	1	\cr
}\right]
\end{equation}
The singular points of the matrix inversion are
 $Q_1=[1,1,1]$, $Q_2=[1,0,-1]$,  $ Q_3=[1,-1,+1]$.

The singular diagram is drawn below
\par\null\hfill
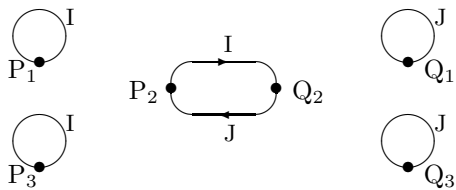
\begin{picture}(100,40)(0,10)
\put(10,30){\circle*{2}}\put(10,35){\circle{10}}
\put(15,37){\small I}\put(4,27){P$_1$}

\put(80,30){\circle*{2}}\put(80,35){\circle{10}}
\put(85,37){\small J}\put(83,27){Q$_1$}

\put(10,10){\circle*{2}}\put(10,15){\circle{10}}
\put(15,17){\small I}\put(4,7){P$_3$}
\put(80,10){\circle*{2}}\put(80,15){\circle{10}}
\put(85,17){\small J}\put(83,7){Q$_3$}

\put(35,25){\circle*{2}}
\put(45,32){\small I}\put(27,23){P$_2$}
\put(55,25){\circle*{2}}\put(45,25){\oval(20,10)}
\put(45,15){\small J}\put(58,23){Q$_2$}
\put(40,30){\vector(1,0){6}}\put(50,20){\vector(-1,0){6}}
\end{picture}
\hfill\null\par
\vskip 1truecm\noindent
It contains  a loop connecting the two points  $P_2$ and $Q_2$. This means that
the points  $P_2=[0,1,0]$ and $Q_2=C^{-1}(P_2)$ are not singular for $IJ$ and
hence it is no surprise that they are non singular for the invariant
$$ \Delta_{AT}= \dsl \frac{y^2-xz}{y(x-z)}.$$
There is a strict inclusion of $S(\Delta) $ in $S(\G)$, since  the points $P_2$
and $Q_2$ are not in $S(\Delta)$.
The corresponding divisors $\Pi_{P_2} =\{ \mbox{ the line } y=0 \}$ (resp
$\Pi_{Q_2} = \{\mbox{  the line } x=z \}$), are indeed curves in the pencil
$\Delta=const$.

\subsection{A finite realization} \label{fm}
It is instructive to consider the group  described by the collineation matrix
\begin{equation}
C_F(q) = \left[
	\matrix{	1&	0&	1	\cr
			1&	q&	-(1+q)	\cr
			1&	0&	-1	\cr
	}\right]
\end{equation}
which is invertible  for $q\neq 0$.
It is  apparent  that $I$ and $J$ share $P_2=[0,1,0]$ as a {\em common\/}
singular point.
Apart from $P_2$,  $I$ admits as singular points $Q_1=[1,1,1]$ and $
Q_3=[q,-(2+q),-q]$.
The singular graph $D_\Gamma$ depicted below contains two  more points
$R=J(Q_3)$ and $S=I(Q_2)$.
This model admits at least {\em two} algebraically independent  invariants,
which can be taken to be
$$\Delta_F^{(1)}={{z^4+x^4}\over{z^2x^2}}$$
$$\Delta_F^{(2)}={{q^2zx(y-z)^2+q[(z^2-xy)^2-z(z-2x)(x^2+y^2)]+2(z^2-xy)^2}
\over{(z-y)^2(z^2+x^2)}}$$
This, together with Bezout's  theorem tells us that the orbit under $\Gamma$ of
any point $p\in \CP_2$ is finite and of order not greater that $16$, i.e $\G$
is {\em finite\/}.
A  direct inspection  shows that $(IJ)^4={\bf 1}$ and  there are additional
invariants, of course not algebraically independent from the two previous ones.

\par\null\hfill
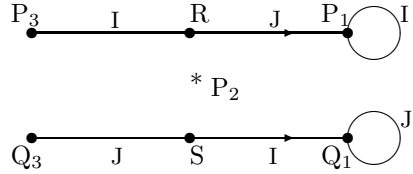
\begin{picture}(90,40)
\put(40,20){$\ast$}\put(44,18){P$_2$}
\put(40,10){\circle*{2}}
\put(40,5){S}
\put(40,30){\circle*{2}}
\put(40,32){R}
\put(10,10){\circle*{2}}\put(6,5){Q$_3$}
\put(10,30){\circle*{2}}\put(6,32){P$_3$}
\put(70,10){\circle*{2}}\put(65,5){Q$_1$}
\put(70,30){\circle*{2}}\put(65,32){P$_1$}
\put(75,10){\circle{10}}\put(75,30){\circle{10}}
\put(80,12){\small J}\put(80,32){\small I}
\put(10,10){\line(1,0){60}}\put(25,5){\small J}
\put(40,10){\vector(1,0){20}}
\put(55,5){\small I}
\put(10,30){\line(1,0){60}}\put(25,31){\small I}
\put(40,30){\vector(1,0){20}}
\put(55,31){\small J}
\end{picture}
\hfill\null\par

\subsection{The symmetric $Z_7$ model}
The symmetric $Z_7$ model~\cite{bmv1} may be   defined by the collineation
matrix
\begin{equation}
C_{Z_7}
=\left[\matrix{
	2	&6	&6	\cr
	2	&{{-1-i\sqrt{7}}}	&{{-1+i\sqrt{7}}} \cr
	2	&{{-1+i\sqrt{7}}}	&{{-1-i\sqrt{7}}}\cr
}\right]
\end{equation}

{}From the point of view of dynamical systems the   model shows chaotic
properties in $\CP_2$.
It has  an {\em infinite\/} singular orbit $S(\Gamma)$.
According to proposition  5.1, it does not admit any invariant, as it is
confirmed by a direct inspection of the orbits.
This model behaves actually like a generic element of the family
(\ref{jaeger}).

\subsection{A finite diagram model (FDM)}
\label{fdm}
Here  $\Gamma$ is  generated by the Hadamard inverse $J$ and
$I= D^{-1} J D$ with $$D=\left[\matrix{
2	&0	&2\cr
1	&1	&-1\cr
-1	&1	&1\cr }\right]$$
\noindent Notice that $D$ is not of the form (\ref{jaeger}).
The singular diagram is finite   but $\G$ does  not have any invariant.
This provides a remarkable example for what  we said in section(\ref{cex}),
i.e. that the finiteness of the singular  orbit is  only a necessary condition
for the existence of a  non--trivial invariant.
\par\null\hfill
\setlength{\unitlength}{.9mm}
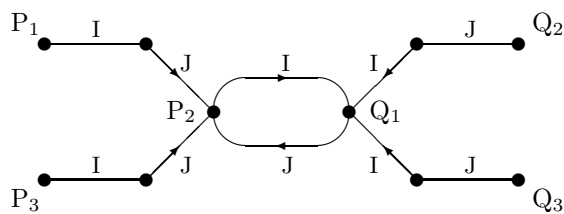
\begin{picture}(85,40)
\put(5,10){\circle*{2}}\put(5,30){\circle*{2}}
\put(20,10){\circle*{2}}\put(20,30){\circle*{2}}
\put(30,20){\circle*{2}}\put(50,20){\circle*{2}}
\put(60,10){\circle*{2}}\put(60,30){\circle*{2}}
\put(75,10){\circle*{2}}\put(75,30){\circle*{2}}
\put(40,20){\oval(20,10)}
\put(35,25){\vector(1,0){6}}\put(45,15){\vector(-1,0){6}}
\put(5,10){\line(1,0){15}}\put(5,30){\line(1,0){15}}
\put(60,10){\line(1,0){15}}\put(60,30){\line(1,0){15}}
\put(20,10){\line(1,1){10}}\put(20,30){\line(1,-1){10}}
\put(50,20){\line(1,1){10}}\put(50,20){\line(1,-1){10}}
\put(20,10){\vector(1,1){5}}\put(20,30){\vector(1,-1){5}}
\put(60,30){\vector(-1,-1){5}}\put(60,10){\vector(-1,1){5}}

\put(23,19){P$_2$}\put(53,19){Q$_1$}
\put(0,32){P$_1$}\put(77,32){Q$_2$}
\put(0,6){P$_3$}\put(77,6){Q$_3$}
\put(12,11){\small I}\put(12,31){\small I}
\put(67,11){\small J}\put(67,31){\small J}
\put(25,26){\small J}\put(25,11){\small J}
\put(53,26){\small I}\put(53,11){\small I}
\put(40,26){\small I}\put(40,11){\small J}
\end{picture}
\hfill\null\par

\section{About complexity}
In this section we will discus the notion of complexity of the groups we have
been considering.

As we said earlier, we may  consider  the group $G$  as generated by
involutions and relations.
If the number $\nu$ of generators is bigger than 2,  $G$ is ``exponentially
big''~\cite{Gr81}.
We will not comment on this here, and limit ourselves to $\nu=2$,  as one would
concentrate on 1--dimensional subgroups of differentiable  groups, and examine
the realizations $\G$.

What we want to point out  is that there is a very diverse behaviour of these
realizations , {\em even in the case $\nu=2$}, manifesting itself in  the
complexity of the  representing transformations. The notion of complexity
we appeal to is a simple form of the  one introduced in~\cite{Ar90}.

Suppose $F$ is a  diffeomorphism of a compact smooth n--manifold $M$.
Let $S_k$ and $R_{l}$ submanifolds of $M$ of dimension respectively $k$ and
$l$.
Let  $S^{m}_k$ be the m$\rm ^{th}$  iterate of $S_k$ by $F$ and $T_{S,R}(m)$
the intersection
$$T_{S,R}(m)=R_l\cap S^{m}_k.$$
Arnol'd \cite{Ar90} defines the {\em complexity\/} $C_{S,R}(m)$
as $$ C_{S,R}(m)=\sum b_p(T_{S,R}(m))$$\noindent the $b_p$'s being the
Betti numbers. He also proves that for a sufficiently generic choice of $S_k$
and $R_l$ the complexity grows at most exponentially with $m$.

The analysis of complexity of plane mappings has already been achieved for the
case of  {\em polynomial and polynomially invertible\/}
transformations~\cite{FrMi89,Ve92}. We are interested in  a wider class of
transformations, as the one described in section(\ref{cp2}) where $\G$ is
generated by two {\em rational \/}  transformations.

Although these transformations  are not diffeomorphisms there is  a natural
measure of the complexity  by  the degree of iterates.

Generically, if the two generating involutions are of degree $u$ and $v$
respectively, the degree of $\varphi=IK$ is $w=uv$,   so that deg
$\varphi^{(k)}=w^k$.
If $\Sigma_1$ and $\Sigma_2$ are two  linear subspaces of $\CP_2$ then
$\varphi^{(k)}(\Sigma_1)$ should   be for generic $\G$ a curve of degree $w^k$,
and
the complexity $C_{\Sigma_1,\Sigma_2}(k)$ would then be  exactly $w^k$.

However,  due to the fact that we are working with projective space, there is a
simple mechanism for the lowering of the degree of the iterates
$\varphi^{(k)}$, for one has  to factorize out common factors from the
expressions of the homogeneous coordinates of  $\varphi^{(k)}$.
This provides a variety of behaviours for the degree $d$ as a function of the
order of iteration $k$, lying between  exponential  growth and periodicity,
with the particular case of polynomial (or polynomially bounded) growth.

The outcome of our analysis is that there is a connection between the existence
of an invariant  and a polynomial (as opposed to exponential)  rate of growth,
as shows Table 1, inferred from the results of the direct calculation of the
first few iterations on the examples of section (\ref{examples}) where the
degree of $\varphi$ is 4.

\begin{table}[bt]
\caption{Behaviour of degrees}\label{table1}
\begin{center}
\begin{tabular}{|c|c|c|}
\hline\bf Model &\bf Growth 		& \bf rank of invariants\\
\hline
Z$_5$ 	& $d(k+1)+d(k-1)-2d(k)$ periodic of period  3		&  one\\
BMV	& $d(k+1)+d(k-1)-2d(k)$ bounded				& one\\
Finite	& $d(k)$ periodic of period 4				& two \\
A--T	& $d(k)= 4 k $						& one\\
FDM	& $d(k)= 4^k$ 	(generic)				& none\\
Z$_7$ 	& $d(k+1)-d(k)=f_{2k}$  ($\{f_k\}$ a Fibonacci sequence)& none \\
\hline
\end{tabular}
\end{center}
\end{table}

A more detailed analysis of these properties, together with the study of other
properties of realizations of Coxeter groups {\it qua\/} dynamical systems such
as finite (periodic) orbits is the matter  of further investigations
(see~\cite{bbb6,bmr1}).

{\bf Acknowledgments} We thank M. Talon, O. Babelon, M. Bellon, J-M. Maillard,
and G. Rollet for a number of stimulating discussions.

\end{document}